\documentclass[aps,prd,twocolumn,superscriptaddress,showkeys]{revtex4-2}
\usepackage{epsfig}
\usepackage{graphicx}
\usepackage{amsmath,amssymb}

\newcommand{\bastar}{\begin{eqnarray*}}
\newcommand{\eastar}{\end{eqnarray*}}
\newskip\humongous \humongous=0pt plus 1000pt minus 1000pt

\newif\ifdtup

\relax
\newcommand{\bea}{\begin{eqnarray}}
\newcommand{\eea}{\end{eqnarray}}
\newcommand{\X}{\vec X}
\newcommand{\W}{\vec W}
\newcommand{\pro}{\partial}
\newcommand{\pd}{\partial}
\newcommand{\n}{\hat n}

\newcommand{\mn}{{\mu\nu}}

\newcommand{\F}{\vec F}

\newcommand{\hF}{\hat F}

\newcommand{\A}{\vec A}
\newcommand{\hA}{\hat A}
\newcommand{\cA}{\cal A}
\newcommand{\cC}{\cal C}
\newcommand{\tA}{\tilde A}
\newcommand{\tC}{\tilde C}

\newcommand{\hD}{\hat D}

\newcommand{\nn}{\nonumber}

\newcommand{\bp}{\bar p}
\begin{document}
\title{Experimental Verification of Two types of 
Gluon Jets in QCD}
\bigskip
\author{Y. M. Cho}
\email{ymcho0416@gmail.com}
\affiliation{School of Physics and Astronomy,
Seoul National University, Seoul 08826, Korea}
\affiliation{Center for Quantum Spacetime, 
Sogang University, Seoul 04107, Korea}  
\author{Pengming Zhang}
\email{zhangpm5@mail.sysu.edu.cn}
\affiliation{School of Physics and Astronomy,
Sun Yat-Sen University, Zhuhai 519082, China}
\author{Li-Ping Zou}
\email{zoulp5@mail.sysu.edu.cn}
\affiliation{Sino-French Institute of Nuclear Engineering and Technology,
Sun Yat-Sen University, Zhuhai 519082, China}

\begin{abstract}
The Abelian decomposition of QCD tells that there are 
two types of gluons, the color neutral neurons and colored 
chromons. We propose to confirm the Abelian decomposition 
testing the existence of two types of gluon jets 
experimentally. We predict that one quarter of the gluon jet
is made of the neurons which has the color factor 3/4 
and the sharpest jet radius and smallest charged particle 
multiplicity, while the three quarters of the gluon jet are 
made of the chromons with the color factor 9/4 which have 
the broadest jet radious (broader than the quark jet). 
Moreover, we argue that the neuron jet has a distinct color 
flow which forms an ideal color dipole, while the quark 
and chromon jets have distorted dipole pattern. To test 
the plausibility of this proposal we suggest to analyse 
the gluon distribution against the jet shape (the sphericity) 
and/or particle multiplicity from the existing gluon jet 
events and look for two distinct peaks in the distribution. 

\end{abstract}
\maketitle

\section{Introduction}

A common misunderstanding on QCD is that the non-Abelian 
color gauge symmetry is so tight that it defines the theory 
almost uniquely, and thus does not allow any simplification. 
This is not true. The Abelian decomposition of QCD tells 
that we can construct the restricted QCD (RCD) which 
inherits the full non-Abelian color gauge symmetry with 
the restricted potential obtained by the Abelian projection. 
This tells that QCD has a non-trivial core, RCD, which 
describes the Abelian sub-dynamics of QCD but has the full 
color gauge symmetry. Moreover, it tells that QCD can be 
viewed as RCD which has the gauge covariant valence gluons 
as the colored source  \cite{prd80,prl81}. This is because 
the Abelian decomposition decomposes the color gauge potential 
to the restricted potential made of the color neutral gluon 
potential, the topological monopole potential, and the gauge 
covariant valence potential which describes the colored gluon 
gauge independently.  

There are ample motivations for the Abelian decomposition. 
Consider the proton made of three quarks. Obviously 
we need the gluons to bind the quarks in the proton. 
However, the quark model tells that the proton has no 
valence gluon. If so, what is the binding gluon which 
bind the quarks in proton, and how do we distinguish 
it from the valence gluon? 

Moreover, the simple group theory tells that the color 
gauge group has the Abelian subgroup generated by 
the diagonal generators, and that the gauge potential 
which corresponds to these generators must be color 
neutral, while the potential which corresponds to 
the off-diagonal generators must carry the color. 
This strongly implies that there are two types of gluon,
the color neutral ones and colored ones. And they 
should behave differently, because they have different 
color charges. If so, how can we distinguish them?   

Another motivation is the color confinement in QCD. 
Two popular proposals for the confinement are the monopole condensation \cite{prl81,nambu} and the Abelian 
dominance \cite{thooft,prd00}. To prove the monopole 
condensation, we first have to separate the monopole 
potential gauge independently. Similarly, to prove 
the Abelian dominance we have to know what is the Abelian 
part and how to separate it. 

The Abelian decomposition tells how to do this. It 
decomposes the non-Abelian gauge potential to two 
parts, the restricted Abelian part which has the full 
non-Abelian gauge symmetry and the gauge covariant 
valence part which describes the colored gluons. 
Moreover, it separates the restricted potential to 
the non-topological Maxwell part which describes 
the colorless binding gluons and the topological 
Dirac part which describes the non-Abelian monopole \cite{prd80,prl81}. 

This has deep consequences. It tells that QCD has two 
types of gluons, the color neutral binding gluons 
(the neurons) and the colored valence gluons (the chromons), 
which play totally different roles. The neurons, just like 
the photon in QED, play the role of the binding gluon. 
On the other hand the chromons, like the quarks, play 
the role of the constituent gluon. This has deep impact 
in hadron spectroscopy, replacing the quark and gluon model 
by the quark and chromon model.  

Moreover, this allows us to decompose the QCD Feynman 
diagrams made of the gluon propagators in terms of the neuron 
and chromon propagators, in such a way that the conservation 
of color is made explicit. 

As importantly, this allows us to prove the Abelian dominance, 
that RCD is responsible for the confinement \cite{thooft,prd00}. 
This is because the chromons, being colored, have to be 
confined. So it can not play any role in the confinement. 
Furthermore, this provides us an ideal platform for us to prove 
the monopole condensation. Indeed, integrating 
out the chromons under the monopole background gauge invariantly, 
we can demonstrate that the true QCD vacuum is given by stable monopole condensation \cite{prd13,epjc19}. 

This makes the experimental verification of the Abelian 
decomposition an urgent issue \cite{uni19}. The prediction 
and subsequent confirmation of the gluon jet was a great 
success of QCD \cite{ellis,gjet}. It proved that QCD 
is indeed the right theory of strong interaction. 
Moreover, it justified the asymptotic freedom and extended 
our understanding of QCD greatly \cite{wil}. Certainly 
the experimental confirmation of the existence of two types 
of gluons will extend our understanding of QCD to a totally 
new level. The purpose of this paper is to show how to do 
this with the existing data on gluon jet. 
    
\section{Abelian Decomposition: A Review}

To show QCD has two types of gluons, consider the SU(2) 
QCD first. Let $(\n_1,\n_2,\n_3=\n)$ be an arbitrary SU(2) basis 
and select $\n$ to be the Abelian direction. Project out 
the restricted potential $\hA_\mu$ which parallelizes 
$\n$ \cite{prd80,prl81}
\begin{gather}
D_\mu \n=(\pd_\mu+g\A_\mu \times) \n=0,  \nn\\
\A_\mu \rightarrow \hA_\mu
=A_\mu \n-\frac1{g} \n \times \pd_\mu \n
=\tA_\mu+\tC_\mu,  \nn\\
\tA_\mu=A_\mu \n,
~~~\tC_\mu=-\frac1{g} \n \times \pd_\mu \n.
\label{ap}
\end{gather}
The restricted potential is made of two parts, 
the non-topological (Maxwellian) $\tA_\mu$ which 
describes the colorless neuron and the topological 
(Diracian) $\tC_\mu$ which describes the non-Abelian 
monopole \cite{prl80,plb82}. Moreover, it has 
the full SU(2) gauge degrees of freedom. 

With this we have
\begin{gather}
\hF_\mn = (F_\mn+ H_\mn) \n=F'_\mn \n, \nn\\
F_\mn = \pd_\mu A_\nu-\pd_\nu A_\mu,  \nn\\
H_\mn = -\frac1g  \n \cdot (\pd_\mu \n\times \pd_\nu \n)
=\pd_\mu C_\nu-\pd_\nu C_\mu,  \nn\\
C_\mu=-\frac1g \n_1 \cdot \pd_\mu \n_2,  \nn\\
F'_\mn=\pd_\mu A'_\nu-\pd_\nu A'_\mu,
~~~A'_\mu=A_\mu+C_\mu.
\label{rf} 
\end{gather}
From this we can construct RCD which has the full 
non-Abelian gauge symmetry,
\begin{gather}
{\cal L}_{RCD} =-\frac14 \hF^2_\mn=-\frac14 F_\mn^2 \nn\\
+\frac1{2g} F_\mn \n \cdot (\pd_\mu \n \times \pd_\nu \n)
-\frac1{4g^2} (\pd_\mu \n \times \pd_\nu \n)^2,
\label{rcd}
\end{gather}
which describes the Abelian sub-dynamics of QCD. 

We can express the full SU(2) gauge field adding 
the gauge covariant colored chromon $\X_\mu$ to 
$\hA_\mu$ \cite{prd80,prl81}
\begin{gather}
\A_\mu = \hA_\mu + \X_\mu,    
~~~~\n \cdot \X_\mu=0,  \nn\\
\F_\mn=\hF_\mn + \hD _\mu \X_\nu 
- \hD_\nu \X_\mu + g\X_\mu \times \X_\nu.
\label{adec}
\end{gather}
With this we recover the full SU(2) QCD 
\begin{gather}
{\cal L}_{QCD} =-\dfrac{1}{4} \hF_\mn^2
-\dfrac{1}{4}(\hD_\mu\X_\nu-\hD_\nu\X_\mu)^2 \nn\\
-\dfrac{g}{2} {\hF}_\mn \cdot (\X_\mu \times \X_\nu)
-\dfrac{g^2}{4} (\X_\mu \times \X_\nu)^2, 
\label{2qcd} 
\end{gather}
This shows that QCD can be viewed as RCD which has 
the chromon as the colored source \cite{prd80,prl81}. 

The Abelian decomposition of SU(3) QCD is more complicated 
but similar \cite{prd13,epjc19}. Let $\n_i~(i=1,2,...,8)$ 
be the orthonormal SU(3) basis and choose $\n_3=\n$ 
and $\n_8=\n'$ to be the two Abelian directions, and 
make the Abelian projection imposing the condition 
\begin{gather}
D_\mu \n=0,~~~~D_\mu \n'=0,   \nn\\
\A_\mu \rightarrow \hA_\mu=A_\mu \n+A_\mu' \n' 
-\frac1g \n\times \pd_\mu \n
-\frac1g \n'\times \pd_\mu \n' \nn\\
=\sum_p \frac23 \hA_\mu^p,~~~(p=1,2,3),    \nn\\
\hA_\mu^p=A_\mu^p \n^p-\frac1g \n^p 
\times \pd_\mu \n^p={\cA}_\mu^p+{\cC}_\mu^p, \nn\\
A_\mu^1=A_\mu,
~~~A_\mu^2=-\frac12 A_\mu
+\frac{\sqrt 3}{2} A_\mu',  \nn\\
A_\mu^3=-\frac12 A_\mu-\frac{\sqrt 3}{2}A_\mu', 
~~~~~\n^1=\n,  \nn\\
\n^2=-\frac12 \n +\frac{\sqrt 3}{2} \n',  
~~~\n^3=-\frac12 \n -\frac{\sqrt 3}{2} \n'.
\label{cp3}
\end{gather}
Notice that, although SU(3) has only two Abelian 
directions, $\hA_\mu$ can be expressed by three SU(2) 
restricted potential $\hA_\mu^i~(i=1,2,3)$ in Weyl 
symmetric way, symmetric under the permutation of 
three SU(2) subgroups (or equivalently permutation of 
three Abelian directions $\n^i$). With this we have 
the Weyl symmetric SU(3) RCD
\begin{gather}
{\cal L}_{RCD} = -\frac14 \hF_\mn^2
=-\sum_p \frac16 (\hF_\mn^p)^2, 
\label{rcd3}
\end{gather}
which has the full SU(3) gauge symmetry. 

\begin{figure}
\includegraphics[height=2.5cm, width=7cm]{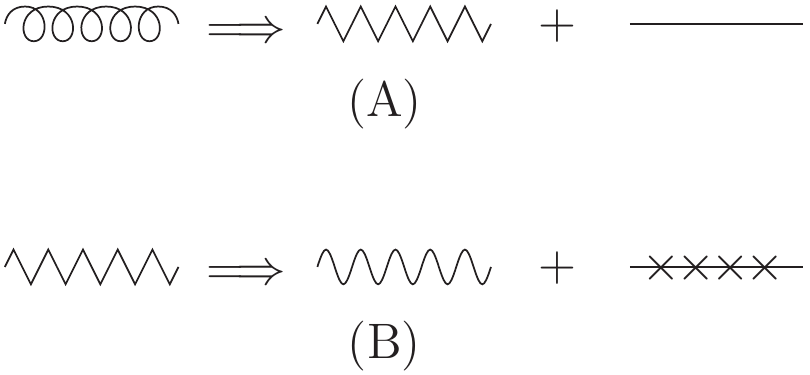}
\caption{\label{cdec} The Abelian decomposition 
of the gauge potential. In (A) it is decomposed to 
the restricted potential (kinked line) and the chromon
(straight line). In (B) the restricted potential is 
further decomposed to the neuron (wiggly line) and 
the monopole (spiked line).}
\end{figure}

Adding the valence part $\X_\mu$ to $\hA_\mu$ we have 
the SU(3) Abelian decomposition,
\begin{gather}
\A_\mu=\hA_\mu+\X_\mu
=\sum_p (\dfrac23 \hA_\mu^p+\W_\mu^p), \nn\\
\F_\mn=\hF_\mn + \hD _\mu \X_\nu 
-\hD_\nu \X_\mu + g\X_\mu \times \X_\nu  \nn\\
=\sum_p \big[\frac23 \hF_\mn^p
+ (\hD_\mu^p \W_\nu^p-\hD_\mu^p \W_\nu^p) \big]  \nn\\
+\sum_{p,q}\W_\mu^p \times \W_\nu^q,    \nn\\
\W_\mu^1= X_\mu^1 \n_1+ X_\mu^2 \n_2,
~~~\W_\mu^2=X_\mu^6 \n_6 + X_\mu^7 \n_7,   \nn\\
\W_\mu^3= X_\mu^4 \n_4  +X_\mu^5 \n_5,
\label{cdec3}
\end{gather}
where $\hD_\mu^p=\pro_\mu+ g \hA_\mu^p \times$. 
Notice that $\X_\mu$ is decomposed to three 
(red, blue, and green) SU(2) chromons 
$(\W_\mu^1,\W_\mu^2,\W_\mu^3)$. Here again 
$\A_\mu$ is expressed in a Weyl symmetric way, 
but unlike $\hA_\mu^i$, the three chromons are 
completely independent.

From this we obtain \cite{prd13,epjc19}
\begin{gather}
{\cal L}_{QCD}=\sum_p \Big\{-\dfrac{1}{6} (\hF_\mn^p)^2 \nn\\
-\frac14 (\hD_\mu^p \W_\nu^p- \hD_\nu^p \W_\mu^p)^2 
-\frac{g}{2} \hF_\mn^p \cdot (\W_\mu^p 
\times \W_\nu^p) \Big\} \nn\\
-\sum_{p,q} \frac{g^2}{4} (\W_\mu^p 
\times \W_\mu^q)^2 \nn\\
-\sum_{p,q,r} \frac{g}2 (\hD_\mu^p \W_\nu^p
- \hD_\nu^p \W_\mu^p) \cdot (\W_\mu^q 
\times \W_\mu^r)  \nn\\
-\sum_{p\ne q} \dfrac{g^2}{4} \Big[(\W_\mu^p 
\times \W_\nu^q)
\cdot (\W_\mu^q \times \W_\nu^p)  \nn\\
+(\W_\mu^p \times \W_\nu^p)\cdot (\W_\mu^q 
\times \W_\nu^q) \Big].
\label{3qcd}
\end{gather}
This is the Abelian decomposition of the Weyl 
symmetric SU(3) QCD.

We can add quarks in the Abelian decomposition,
\bea
&{\cal L}_{q} =\bar \Psi (i\gamma^\mu D_\mu-m) \Psi \nn\\
&= \bar \Psi (i\gamma^\mu \hD_\mu-m) \Psi 
+\dfrac{g}{2} \X_\mu \cdot \bar \Psi (\gamma^\mu \vec t) \Psi  \nn\\
&=\sum_p \Big[\bar \Psi^p (i\gamma^\mu \hD_\mu^p-m) \Psi^p
+\dfrac{g}{2} \W_\mu^p \cdot \bar \Psi^p
(\gamma^\mu \vec \tau^p) \Psi^p \Big], \nn\\
&\hD_\mu = \pro_\mu +\dfrac{g}{2i} {\vec t}\cdot \hA_\mu,
~\hD_\mu^p=\pro_\mu
+\dfrac{g}{2i} {\vec \tau^p}\cdot \hA_\mu^p,
\label{qlag}
\eea
where $m$ is the mass, $p$ denotes the color of the quarks, 
and $\Psi^p$ represents the three SU(2) quark doublets 
(i.e., $(r,b)$, $(b,g)$, and $(g,r)$ doublets) of the $(r,b,g)$ 
quark triplet. Notice that here we have suppressed the flavour 
degrees \cite{prd13,epjc19}. 

The Abelian decomposition is expressed graphically in 
Fig. \ref{cdec}. Although the decomposition does not c
hange QCD, it reveals the important hidden structures 
of QCD. In particular, it shows the existence of two 
types of gluons, the neuron and chromon. In the literature 
the Abelian decomposition is known as the Cho decomposition, Cho-Duan-Ge (CDG) decomposition, or Cho-Faddeev-Niemi (CFN) 
decomposition \cite{fadd,shab,zucc,kondor}.

\begin{figure}
\includegraphics[height=4.5cm, width=7cm]{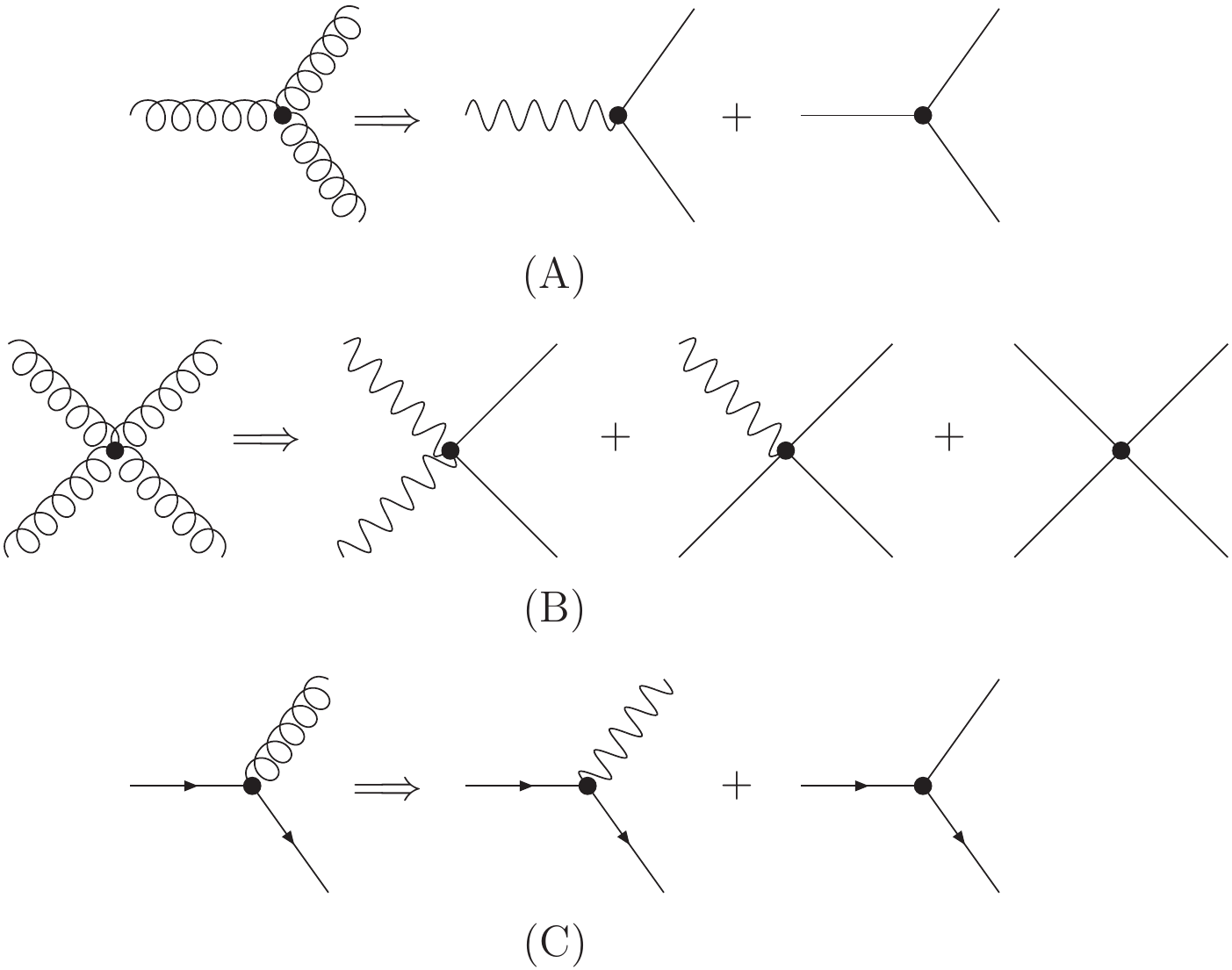}
\caption{\label{3qcdint} The decomposition of 
the Feynman diagrams in SU(3) QCD. In (A) and (B)
the three-point and four-point gluon vertices are 
decomposed, and in (C) the quark-gluon vertices 
are decomposed. Notice that the monopole does not 
appear in the Feynman diagrams because it describes 
topological degree, not dynamical degree.}
\end{figure}

The Abelian decomposition has deep implications. In 
the perturbative regime this tells that the Feynman diagram 
can be decomposed in such a way that the color conservation 
is explicit. This is graphically shown in Fig. \ref{3qcdint}. 
Notice that here the monopole does not appear in the Feynman 
diagram, because it is not a dynamical degree. Moreover, 
it makes the condensation so that it has no role in 
the perturbative regime. 

A remarkable feature of the decomposition of the Feynman 
diagram is that the color conservation is explicit in the decomposition. There are no three-point vertex made of two 
or three neuron legs, and no four-point vertex made of 
three or four neuron legs. This is because the color 
conservation forbids them.   

\begin{figure}
\includegraphics[height=4.5cm, width=6cm]{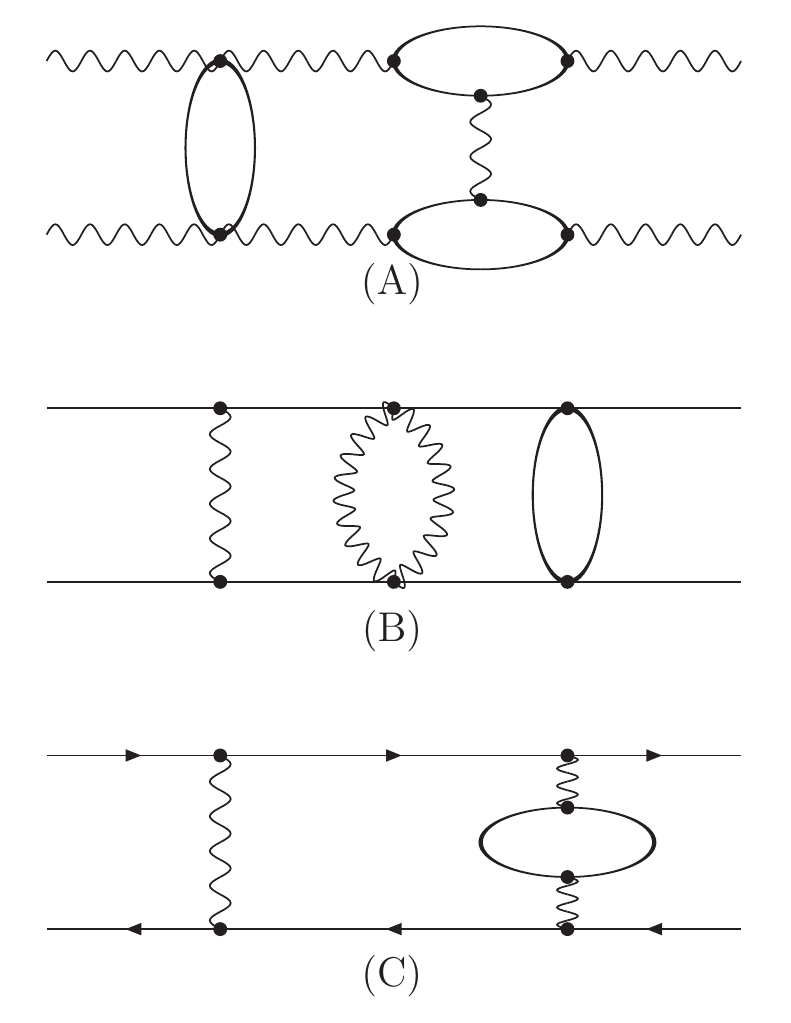}
\caption{\label{bind} The possible Feynman diagrams 
of the neurons and chromons. Two neuron interaction 
is shown in (A), chromon-antichromon interaction 
is shown in (B), and quark-antiquark interaction is 
shown in (C).}
\end{figure}

As importantly this shows that neurons and chromons 
play totally different roles. The neuron, just like 
the photon in QED, provides the binding. But the chromons, 
just like the quarks, become the colored source. This 
could be shown graphically by the Feynman diagrams of 
two neuron binding and chromon-antichromon binding shown 
in Fig. \ref{bind}. 

Notice that the two neuron binding in (A) looks totally 
different from the other two bindings. The leading order 
of this binding is of the order of $O(g^4)$, while 
the leading order ofthe other two in (B) and (C) is 
of the order of $O(g^2)$. Moreover, the neuron binding 
looks very much like two photon binding in QED, while 
the chromon-antichromon binding look just like 
the quark-antiquark binding in QCD. 

This strongly implies that the neurons may not be viewed 
as the constituent of hadrons. In contrast, the chromon 
binding strongly implies that they, just like the quarks, 
become the constituent of hadrons. This changes the quark 
and gluon model to the quark and chromon model, which 
provides a new picture of hadrons \cite{prd15,prd18}. 

\section{Monopole Condensation}

In the non-perturbative regime, the Abelian decomposition 
allows us to proves the Abelian dominance, or more precisely 
the monopole dominance, that the monopole plays the central 
role in color confinement. The logic for the monopole dominance 
is the following. First, the chromons (being colored) are 
destined to be confined, so that they can not be the confining 
agent. This is because the prisoners (who are confined in jail) 
can not play the role of the jailer (who confines the prisoners). 

This means that only the restricted potential plays the important 
role in the confinement, which proves the Abelian dominance 
that 'tHooft proposed \cite{thooft,prd80,prl81}. Indeed, 
theoretically we can show rigorously that only the restricted potential contributes to the Wilson loop integral which 
produces the area law for the confinenent \cite{prd00}. 

But the the restricted potential is made of two parts, 
the non-topological neuron and topological monopole. 
On the other hand Fig. \ref{bind} tells that the neuron 
in QCD plays the role of the photon in QED. This implies that 
it can not play any role in the confinement. This tells that 
only the monopole can confine the color \cite{prd13,epjc19}. 

\begin{figure}
\psfig{figure=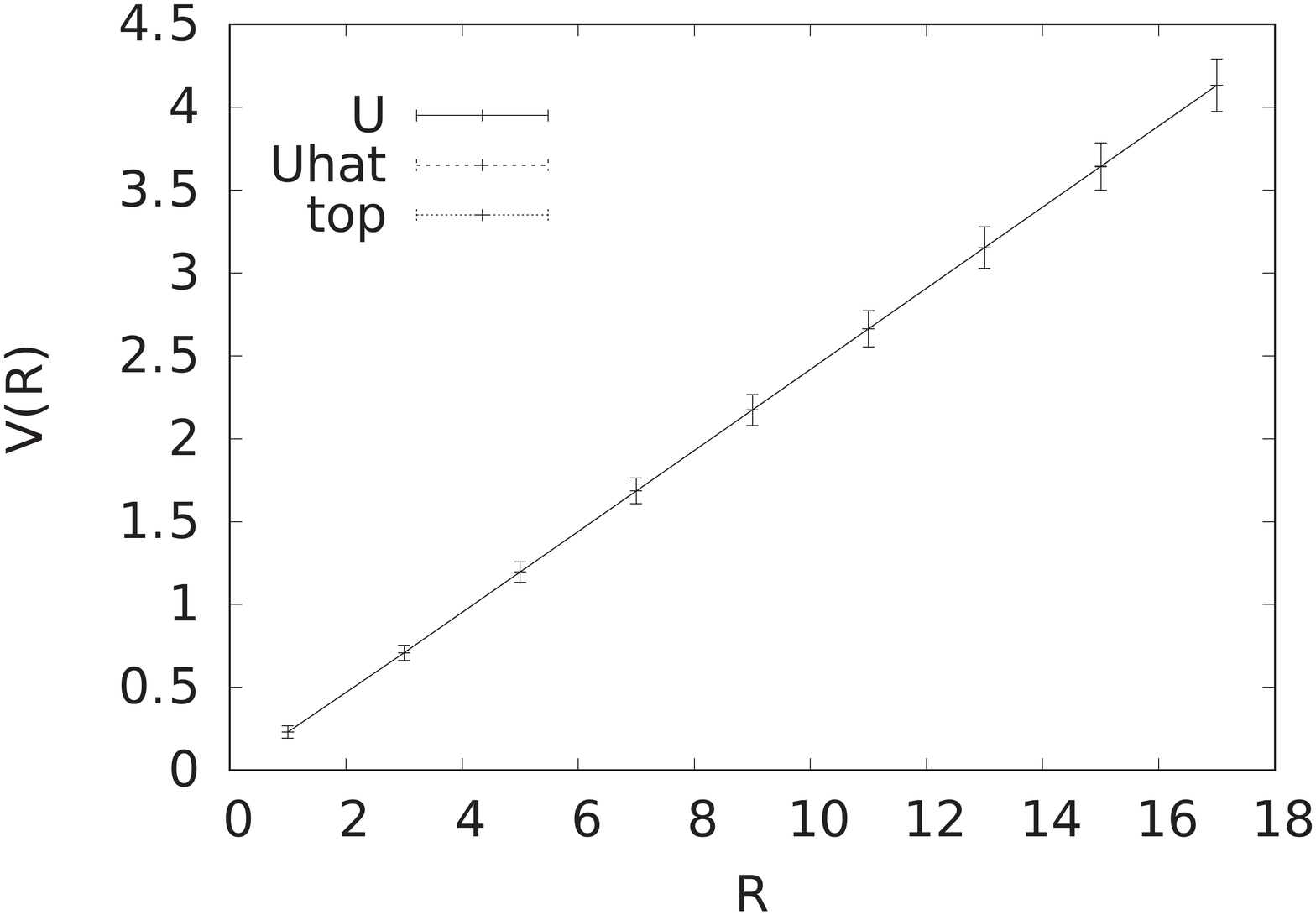, height=4.5cm, width=8cm}
\caption{\label{cundy} The SU(3) lattice QCD calculation 
which establishes the monopole dominance in the confining 
force in Wilson loop. Here the confining forces shown in full, 
dashed, and dotted lines are obtained with the full potential, 
the Abelian potential, and the monopole potential, 
respectively.}
\end{figure}

Indeed, with the decomposition we can regorously prove 
that only the restricted potential contributes to 
the area law in the Wilson loop integral \cite{prd00}. 
This is the Abelian dominance. This can be confirmed 
numerically in lattice QCD. Implementing the Abelian decomposition on the lattice we can calculate the Wilson 
loop integral with the full potential, the restricted 
potential, and the monopole potential separately, 
and show that all three potentials produce exactly 
the same linear confining force \cite{cundy,kondo}. 
The lattice result for SU(3) QCD is shown in 
Fig. \ref{cundy}. This tells that the monopole plays 
the crucial role in the confinement. 

On the other hand, this lattice result does not tell 
how the monopole can confine the color. To show that 
the monopole confines the color by dual Meissner effect 
as Nambu and Mandelstam conjectured, we have to demonstrate 
the monopole condensation in QCD. The Abelian decomposition 
provides us an ideal platform to calculate the QCD 
effective potential and prove the monopole condensation 
gauge independently. This is because it puts QCD to 
the background field formalism, since we can treat 
the restricted part and the valence part as the slow varying 
classical background and the fast moving quantum 
fluctuation \cite{prd01,dewitt}. 

Indeed, choosing the monopole potential as the background 
and integrating out the chromons in (\ref{3qcd}) gauge 
invariantly, we can calculate the QCD effective potential 
which has the Weyl symmetric unique minimum described by 
the monopole condensation. This prove that the true QCD 
vacuum is given by the gauge invariant monopole 
condensation \cite{prd13,epjc19}. 

\section{Neuron Jet and Chromon Jet}

The Abelian decomposition is not just a theoretical 
proposition. There are many ways to test it experimentally. 
For example, we can test it by showing the quark 
and chromon model describes the correct hadron 
spectrum \cite{prd15,prd18}. Or we can test it by 
demonstrating that the monopole condensation does describe 
the true QCD vacuum \cite{prd13,epjc19}. But these are 
indirect tests. If QCD really has two types of gluons, 
we should be able to confirm this directly by experiment. 

During the last twenty years there has been huge progress 
on jet structure in QCD. New ways to tag different jets 
have been developed \cite{je1,je2,je3}.  Moreover, 
new features of the quark and gluon jet substructures 
have been known \cite{jt1,jt2,jt3,jt4}. Now, we argue 
that these progresses could allow us to confirm 
the existence of two types of gluon jets experimentally. 

Early experiments which established the existence of 
the gluon are based on planar three jet events made of 
two quark jets and one gluon jet \cite{gjet}. 
So the problem here is to devide the gluon jets to neuron 
jet and chromon jet. For this we have two questions. 
First, how many of them are the neuron jet? Second, 
how can we differentiate the neuron jet from the chromon 
jet to identify the existence of the neuorn jet?

The first question is easy. Since two of the eight gluons 
are neurons, one quarter of the gluon jet should be the neuron 
jet and three quarters of them should be the chromon jet. 
The difficult problem is the second question. Actually 
the problem here is not just how to separate the neuron jet 
from the chromon jet. Since we have the quark jet as well 
in QCD, we have to tell how to tag all three jets, the neuron 
jet, the chromon jet, and the quark jet, separately. So we 
have to know the difference of each jet from the other two. 

\begin{figure}
\psfig{figure=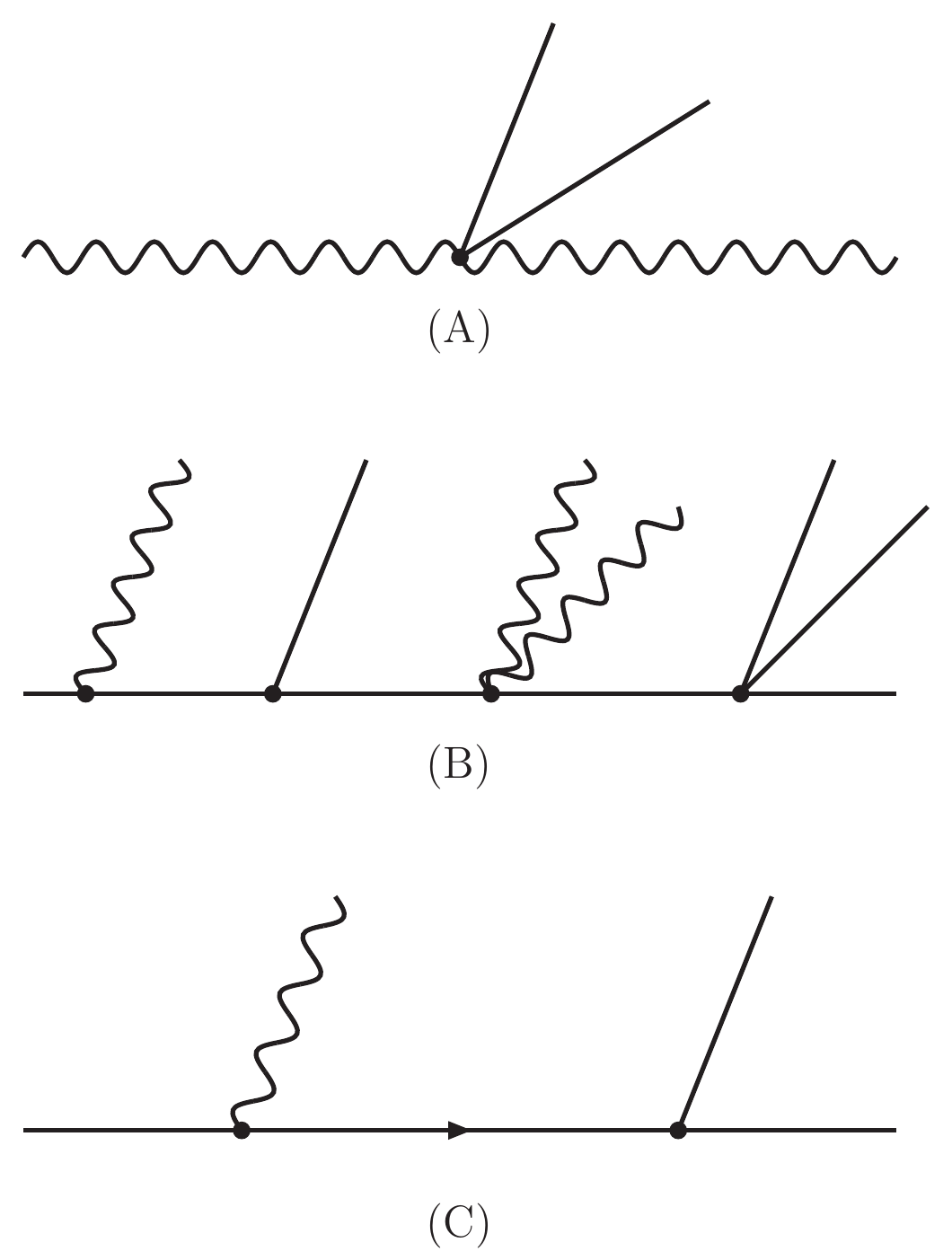, height=4.5cm, width=7.5cm}
\caption{\label{jet} The parton shower (the soft gluon 
radiation) of hard partons. The neuron jet shown in (A) 
is qualitatively different from the chromon jet and 
the quark jet shown in (B) and (C), while the chromon 
and quark jets are similar.}
\end{figure}

To tell the difference it is important to remember that 
the gluons and quarks emitted in the p-p collisions evolve 
into hadron jets in two steps, the soft gluon radiation 
of the hard partons described by the perturbative process 
and the hadronization described by the non-perturbative 
process. The hadronization in the second step is basically 
the same in all three jets. The difference comes from 
the parton shower (the soft gluon radiation) of the hard 
partons in the first step \cite{uni19}. This is shown in 
Fig. \ref{jet} in the first order interaction. 

Clearly the neuron jet (the soft gluon radiation of hard 
neuron) shown in (A) has only two chromon radiation with 
no other soft gluon radiation which exists in both the chromon 
jet (B) and the quark jet (C). But the chromon and quark 
jets have the neuron and chromon radiations, and qualitatively 
look similar. The only difference is that the chromon jet 
has four point vertex, while the quark jet has only three 
point vertex. As importantly, the leading order of the soft 
gluon radiation of neuron shown in (A) is of $O(g^2)$, 
while that of the chromon and quark shown in (B) and (C) 
are of the order of $O(g)$. This is a direct consequence 
of the decomposition of the Feynman diagram shown in 
Fig. \ref{3qcdint}. 

This tells that the neuron jet is almost like the photon 
jet, which is fundamentally different from the chromon 
and quark jets. This strongly suggests that the neuron 
jet must have different jet shape, different from the chromon 
and quark jets. 

Intuitively, we could imagine that the neuron jet is sharp 
with relatively small jet radius compared to the chromon jet. 
This is because the neuron has only chromon-antichromon 
radiation, while the chromon has three more soft gluon 
radiations (as well as the chromon-antichromon radiation).
And this could only broaden the jet. This is obvious from 
Fig. \ref{jet}. This strongly implies that the neuron jet 
can not be broader than the chromon jet. But, of course, 
to find how much sharp the neuron jet is, we definitely 
need the numerical simulation. 

Moreover, this also strongly implies that the neuron 
jet should have different (charged) particle multiplicity, 
considerably smaller than that of the quark and/or chromon 
jets. This must also be clear from Fig. \ref{jet}, which 
shows that the neuron has weaker (of the order $O(g^2)$)
soft gluon radiation than the chromon and quark.  

Another important feature of the neuron jet is that it
has different color flow. Clearly the chromons and 
quarks carry color charge, but the neurons are color 
neutral. So the neuron jet must have different color flow. 
In fact, Fig. \ref{jet} tells that the color flow of 
the neuron jet generates an ideal color dipole pattern, 
but the other two jets have distorted dipole pattern. 

The above observations show that the neuron jet must 
be quantitatively different from the chromon and quark 
jets, and that we should be able to confirm this 
experimentally. To quantify the differences, of course, 
we need more serious theoretical calculations and numerical simulations. For instance, we need to implement the Abelian decomposition in the existing Pythia and FastJet programs, 
and find how the numerical simulations predict the differences 
between the neuron and chromon jets. 

For this we have to know the color factors of the neuron 
and chromon jets, one of the most important quantities 
that determines the characters of the jet. It has been 
well known that the parton shower (the soft gluon radiation) 
of the quark and gluon jets are proportional to their color 
factors $C_F=4/3$ and $C_A=3$ in the eikonal approximation, 
and that the quark/gluon tagging performance crucially 
depends on their ratio $C_A/C_F$. This means that it is 
important to know the neuron and chromon color factors.  

One might think that the neurons have no color factor, 
but this is not so. Although they are color neutral, they 
are not color singlet. So they have finite color factor. 
But at the moment it appears unclear if one can calculate 
the neuron and chromon color factors from the first 
principle, because the color gauge symmetry is replaced to 
the 24-element color reflection symmetry after the Abelian decomposition \cite{prd80,prl81,prd13,epjc19}. On the other 
hand, from the fact that the gluon color factor is given 
by the trace of the quadratic Casimir invariant made of 
the eight gluon generators, we could assume the neuron 
color factor to be the trace of the quadratic Casimir 
invariant corresponding to the neuron generators. In this 
case we can easily calculate the neuron and chromon color 
factors. Since each of the eight gluon generators contributes 
equally to the gluon color factor, we can deduce the neuron 
color factors to be 3/4, one quarter of the gluon color 
factor 3. By the same reason we can say that the chromon 
color factor must be 9/4. A more intuitive way to understand 
this comes from the simple democracy of the gauge interaction. 
Since the neurons constitute one quarter of eight gluons 
their color factor must be one quarter of the gluon color 
factor 3, that is 3/4.  

According to the above reasoning  the color factor ratio 
of the quark, chromon, and neuron jets should be 
$C_q:C_c:C_n=4/3:9/4:3/4\simeq 1.78:3:1$, since the quark 
color factor is given by $C_F=4/3$. If this is so, 
the recent experiments which separated the quark jet from 
the gluon jet based on the color factor ratio $C_A/C_F=9/4$ 
need to be completely re-analysed \cite{je1,je2,je3}. 

In this respect we notice two interesting reports which 
could support the above interpretation. The re-analysis 
of DELPHI $e^+e^-$ three jet data at LEP strongly 
indicates that actual $C_A/C_F$ could be around 1.74, 
much less than the popular value 2.25 but close to 
our prediction $C_c/C_q=1.69$ \cite{lep,eje3}. Moreover, 
the $p \bp$ D{\O} jets experiment at Fermilab Tevatron 
shows that the quark to gluon jets particle multiplicity 
ratio is around 1.84, again close to our prediction 
1.69 \cite{do}. They could be an indication that 
the observed gluon jets are indeed the chromon jets.    

If this is true, one might ask what are the gluon jets 
identified by ATLAS and CMS. Probably they are the chromon 
jets, because the chromon jet has the characteristics of 
the known gluon jet. This is evident from Fig. \ref{jet}. 
Perhaps a more interesting question is why they have not 
found the neuron jet. There could be two explanations. 
First, they have not searched for the neuron jet yet, 
because they had no motivation to do that. Or they might 
have misidentified some of the neuron jets as the quark 
jet. This is because the color factor of neuron and quark
jets are not much different. This tells that we need 
a more careful analysis of quark and gluon jets. 

A simple way way to search for the neuron jet is to 
concentrate on the gluon jet production process like 
$H \rightarrow gg$ and to try to separate the neuron 
jet from the chromon jet. In an idealized setup with 
$e^+e^-$ colliion we can have the quark and gluon jets 
separately \cite{jt4},
\begin{gather}
quark~jet: ~~e^+e^- \rightarrow \gamma/Z \rightarrow 
u\bar u, d \bar d, s \bar s,  \nn\\
gluon~jet: ~~e^+e^- \rightarrow H \rightarrow gg,
\label{qgjet1}
\end{gather}
when we focus on the light quarks. Similarly, in the $pp$ 
collision we have \cite{jt4}
\begin{gather}
quark~enriched~jet: ~~pp \rightarrow Z + jet,  \nn\\
gluon~enriched~jet: ~~pp \rightarrow dijet.
\label{qgjet2}
\end{gather}
In this case we could forget about the quark jet and 
concentrate on the gluon (enriched) jet, and try to separate 
the neuron jet from the chromon jet. In principle this 
could be simpler for two reasons. First, the parton shower
(the soft gluon radiation) of the neuron and chromon jets  
shown in Fig. \ref{jet} is totally different. Second, 
the ratio of the color factors of the neuron and chromon 
jets becomes bigger $C_c/C_n=3$. This strongly implies that 
about one quarter of the gluon (enriched) jet must be 
the neuron jet, which has sharper shape than the other 
and has much less charged particle multiplicity and 
ideal color dipole pattern.

Independent of the details on the differences between 
the neuron and chromon jets, however, we like to emphasize 
that in principle there is a simple and straightforward way 
to confirm the existence of two types of gluon jets from 
the existing jet data. This is because, independent of 
what is the neuorn jet and what is the chromon jet, 
one quarter of the existing gluon jets should actually 
have different jet shape (the sphericity), particle 
multiplicity, and color dipole pattern. If this is so, 
we can simply plot the jet shape (the solid angle) 
and/or charged particle multiplicity of all existing 
gluon jets, and find the gluon jet distribution made 
of two peaks populated by one quarter and three quarters 
of the jets. 

\begin{figure}
\psfig{figure=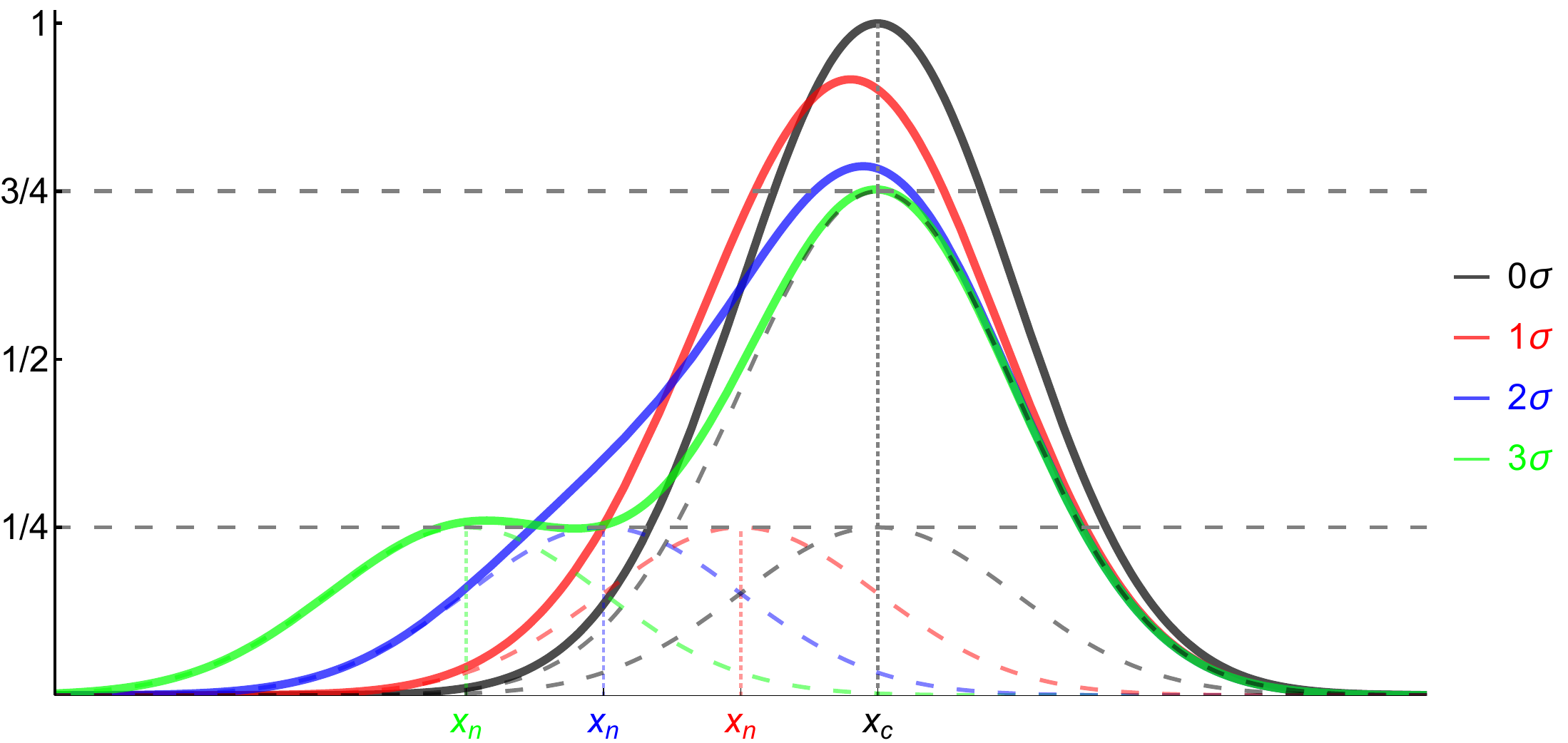, height=4.8cm, width=7cm}
\caption{\label{gjd} The expected gluon jet distribution 
against the jet shape (the sphericity) and/or the particle 
multiplicity. Here we have assmed that the distribution 
is Gaussian, and plotted the overall gluon distribution 
when the distance between the chromon and neuron peaks 
becomes 1, 2, and 3 standard deviations. Notice that, 
when the chromon and neuron jets have the same gluon 
distribution we have the black curve, which represents 
the well known gluon distribution we have in the popular (conventional) QCD.}
\end{figure}

The expected gluon jet distribution is shown in Fig. \ref{gjd}. 
In this figure we have assumed that the distribution of 
neuron and chromon jets are Gaussian, and plot the expected 
gluon jet distribution when the neuron peak is 1, 2, and
3 standard deviations away from the chromon peak. Notice 
that the shape of the overall gluon jet distribution 
crucially depends on the distance between the two peaks. 
For instance, when the distance between the two peaks 
becomes three standard deviation or more, the two peaks 
are well separated. But when the distance becomes two standard deviation or less, the neuron peak is submerged completely 
under the chromon peak as shown in Fig. \ref{gjd}. This 
strongly indicates that the neuron jet could easily be 
left unnoticed in the gluon jet anaysis, when the distance 
between the neuron and chromon peaks becomes less than 
two standard deviation. The exact size (the solid angle) 
and particle multiplicity (number of particles) of the neuron 
and chromon jets could be predicted implementing the Abelian decomposition in the Pythia and FastJet. 

But independent of the details the characteristic feature 
of the above gluon jet distribution is that it is asymmetric 
(tilted) against the peak axis. For example, for the green 
curve (i.e., when the neuron jet is three standard deviation 
apart), the left part of the peak axis has $1.64$ times more 
(64 \% more) gluon jets than the right part, and for the blue 
curve the left part has $1.42$ times more (42 \% more)  
gluon jets than the right part. But for the red curve 
(when the distance between the two peaks becomes one standard deviation) the left part has only $1.06$ times more (6 \% more) 
gluon jets than the right part, so that the asymmetry 
becomes hard to detect but survives. 

This should be contrasted with the black curve, which 
represents the gluon jet distribution when the neuron 
and chromon jets have identical particle multiplicity 
and jet shape. This, of course, is the well known gluon 
distribution we have in the conventional QCD (without 
the Abelian decomposition). In this case the asymmetry 
disappears completely. In other words, when the neuron 
and chromon jets have different jet shape and/or particle multiplicity, the gluon distribution becomes asymmetric. 
This reflection asymmetry (tilt) of the gluon distribution 
against the peak axis is a strong indication that there 
could be two gluon jets. This is the most important 
qualitative feature of the above analysis. 

{\it This tells that, without trying to separate the neuron 
jet from the chromon jet, we could tell the existence 
of the neuron jet in the gluon jet distribution checking 
if the gluon distribution is asymmetric against the peak 
axis or not, from the existing gluon jet data.} This 
reflection asymmetry could be an unmistakable indication 
that there are indeed two types of gluons, the neuron 
and chromon.

A straightforward way to do this is to use the existing 
2071 gluon jet events of ALEPH data coming from 
$e \bar e \rightarrow Z \rightarrow b \bar b g$ three jet 
events \cite{aleph}, find out the distribution of the gluon 
jet on the sphericity and/or particle multiplicity, and 
see if the gluon jet has the predicted distribution shown 
in Fig. \ref{gjd}. In principle this could be done 
without much difficulty. 

One advantage in searching for the neuron jet is that 
we do not need any new experiment. LHC produces 
billions of hadron jets in a second, and ATLAS and CMS 
have already filed up huge data on jets. Moreover, 
DESY, LEP, ALEPH, and Tevatron have old data on three jet 
events (the gluon jets) which we could use to confirm 
the existence of the neuron jet. Here again the simple 
number counting strongly suggests that one quarter of 
the gluon jets coming from the three jets events could 
actually be the neuron jets which do not fit to 
the conventional gluon jet category. 

\section{Discussion}

The gauge potential of QCD is thought to represent 
the gluons, but the role of gluons in QCD has been 
confusing. On the one hand they are supposed to 
provide the binding of the (colored) quarks. But 
at the same time they are supposed to play the role of 
the constituent of hadrons, because they are colored. 
The Abelian decomposition tells that there are two 
types of gluons, the binding gluons and the valence 
gluons, that play different roles.    

As we have emphasized, the fact that there should be 
two types of gluons is evident from the simple group 
theory. Group theory tells that two of eight gluons 
must be color neutral. The question is how to separate 
the neutral gluons gauge independently. The Abelian 
decomposition does the job \cite{prd80,prl81}. As importantly, 
it tells that the Abelian (restricted) potential contains 
not only the neurons but also the topological monopole 
part which plays the crucial role to retain the full 
non-Abelian gauge degrees of freedom to the restricted 
potential. 

This clarifies the role of gauge potential in QCD.
The neurons and monopole potential together bind 
and confine the colored objects with the monopole 
condensation \cite{prd13,epjc19}, but the chromons play 
the role of constituent of hadrons \cite{prd15,prd18}. 

In this paper we have argued that the existence of 
two types of gluon could actually be confirmed by
experiment, and proposed intuitive idea to verify this. 
Of course, to verify this experimentally we need more 
concrete predictions and numerical simulations. But this 
is beyond the scope of this paper, because our aim here 
is to present the theoretical foundation why QCD must have 
two types of gluon, discuss the experimental plausibility 
of neuron and chromon jet tagging, and suggest basic idea 
how we can actually do this without ambiguity. 

In fact experimentally the quark/gluon tagging is a very 
complicated and ongoing issue which is not completely 
settled yet. It is well known that the gluon tagging has 
many unclear and unresolved problems \cite{jt3,jt4}. 
In particular, the success rate of the gluon jet 
identification is known to be at best 70\%. We believe 
that this could be, at least partly, due to the existence 
of two types of gluon jet. In fact we interpret that this 
strongly implies that about 30\% could actually be 
different jet, i.e., the neuron jet. To test the plausibility 
of this suggestion we propose two step analysis. First,
we could look at the existing gluon jet data and check 
if there are events which do not fit to the well known 
predicted characters of the gluon jet, in particular 
the jets which have sharper radius. Second, if we confirm 
this, we could try to identify them as the neuron jet.

Our prediction tells that about one quarter of the gluon 
jets should actually be the neuron jets which have sharper 
jet shape, less charged perticle multiplicity, and ideal 
color dipole pattern. This could be done without much 
difficulty. After we confirm this, we could try to do 
the neuron and chromon jets tagging. A nice thing about 
this proposal is that we do not need any new experiment. 
All that we have to do is to re-analyse the existing old 
data at LHC. 

To do that, however, we have to have quantitative predictions
based on the numerical simulations on the characteristic 
features of the neuron and chromon jets. For this we 
have to modify the existing Pythia and/or FastJet programs implementing the Abelian decomposition, and have new 
Monte Carlo simulations on the neuron and chromon jets 
based on the soft gluon radiations shown in Fig. \ref{jet} 
and the new color factors of the neuron and chromon jets.
On this the recent machine learning algorithm could also 
be very useful for us to find the characteristic features 
of the neuron and chromon jets \cite{jt3,jt4}. But this 
could take time.

What one can do immediately is trying to confirm 
the existence of two types of gluon jet based on the gluon distribution against the jet shape (the sphericity) 
and/or particle multiplicity shown in Fig. \ref{gjd}, 
using the existing ALEPH data \cite{aleph}. Currently 
the Korean CMS team is working on this, and have obtained
some encouraging positive signals which support the existence 
of two types of gluons \cite{cho}. The confirmation of  
the two types of gluons could be a giant step forward that 
will help us to identify the existence of the neuron and 
chromon jets successfully. 

{\bf Note Added in Proof:} Recently it is suggested 
by the MIT high energy experimental group that the anormaly 
in gluon distribution observed in heavy ion collision 
at LHC could be explained by the existence of two types 
of gluon \cite{mit}. This is probably because neuron 
undergoes less gluon quenching than the chromon 
in the collision, and thus could escape the quark-gluon 
plasma fireball more easily. One of us (YMC) thanks 
Yen-Jie Lee for the constructive discussion.       

{\bf Acknowledgements}

~~~The work is supported in part by National Research 
Foundation of Korea funded by the Ministry of Science and 
Technology (Grant 2022-R1A2C1006999), National Natural Science 
Foundation of China (Grants 11975320 and 12175320), and 
by the Center for Quantum Spacetime, Sogang University.

\end{document}